\documentclass[11pt,superscriptaddress,aps,prd]{revtex4-2}

\everymath{\displaystyle}
\usepackage{graphicx}

\usepackage{amsmath,amssymb,mathrsfs}

\newcommand{\bea}{\begin{eqnarray}}
\newcommand{\eea}{\end{eqnarray}}

\begin{document}

\title{On the Thermodynamics of Gravitational Radiation}

\author{S. C. Ulhoa}\email[]{sc.ulhoa@gmail.com}
\affiliation{Instituto de F\'isica, Universidade de Bras\'ilia, 70910-900, Bras\'ilia, DF, Brazil} \affiliation{Canadian Quantum Research Center,\\ 
204-3002 32 Ave Vernon, BC V1T 2L7  Canada} 

\author{F. L. Carneiro}
\email{fernandolessa45@gmail.com}
\affiliation{Universidade Federal do Norte do Tocantins, Aragua\'ina, TO, Brazil}

\author{J. W. Maluf}
\email{jwmaluf@gmail.com}
\affiliation{Instituto de F\'isica, Universidade de Bras\'ilia, 70910-900, Bras\'ilia, DF, Brazil}

\begin{abstract}

This article deals with the thermodynamics of gravitational radiation arising from the Bondi-Sachs space-time. The equation of state found allows us to conclude that the dependence of the energy density on the temperature is a quadratic power of the latter. Such a conclusion is possible once the consequences of the first law of thermodynamics are analyzed. Then, in analogy to electromagnetic radiation, the same approach as used by Planck to obtain the quantum of energy of the gravitational radiation is proposed. An energy for the graviton proportional to the cubic frequency is found. The graviton is here understood as the quantum of gravitational energy.

\end{abstract}
\maketitle

\date{\today}
\section{Introduction} \label{sec.1}

In 1884 Boltzmann derived Stefan's law, in what would come to be known as the Stefan-Boltzmann law~\cite{boltzmann}.  This law stated that the energy density of electromagnetic radiation was proportional to the fourth power of the temperature. The Maxwell's equations alone do not allow one to conclude that electromagnetic radiation is thermal, since such equations do not depend on temperature. They however establish an equation of state for the field, that is, the pressure is equal to one third of the energy. Boltzmann only considered the thermodynamic consequences of the electromagnetic field equation of state to demonstrate the dependence of energy density on temperature. In his seminal paper published in 1901 \cite{planck}, Planck both showed how to obtain the constant of proportionality of the Stefan-Boltzmann law and introduced the Quanta theory. He considered that blackbody radiation could be described by a collection of harmonic oscillators whose energy was discrete.  Each quantum of energy was proportional to the frequency of the radiation. The quantum of energy of electromagnetic radiation became known as the photon and the constant of proportionality as Planck's constant, $h$.  That constant has proved fundamental to every quantum process since then.

The gravitational field has several similarities with the electromagnetic field, for instance the Einstein equations also do not explicitly depend on the temperature.  So there would be no reason to expect any thermal radiation in the framework of such a description of the gravitational field. On the other hand, the very definition of gravitational energy, in the metric formulation of the field, is quite controversial.  In this context, no definition achieves the properties expected of such a quantity.  Namely, the energy must be independent of the coordinate system and it must be sensitive to the choice of the reference frame. There is, on the other hand, a dynamically equivalent description of general relativity, known as Teleparallelism Equivalent to General Relativity (TEGR), in which quantities such as energy, momentum, and angular momentum are well defined \cite{maluf}. Armed with such a description of the gravitational field, one can consider the thermodynamic consequences of a specific equation of state.  In particular, the gravitational radiation is of special interest because, like the electromagnetic field, it can reveal quantum aspects of the physical system.

In the same way that Maxwell's theory allows us to establish an explicit relationship between the energy density $\epsilon$ and the electromagnetic radiation pressure $p$, and consequently, derive a relation between energy density and temperature based on fundamental thermodynamics, our aim is to establish a thermodynamic description of gravitational radiation. In order to describe the gravitational radiation, we consider the Bondi-Sachs spacetime, which represents a source emitting gravitational waves. The energy of this spacetime consists of a rapidly decaying mass aspect and another term that decays more slowly. We focus on the metric in the limit of being far from the source, but not at infinity. In this limit, the contribution from the source (or mass aspect) vanishes, while the other term remains finite. Consequently, we identify this remaining term as the energy of the gravitational radiation. By evaluating the gravitational pressure in the same limit, we can compare the two expressions and obtain an equation of state for gravitational radiation. From this point, we pursue Planck's idea and, through the application of the first law of thermodynamics, establish a relationship between the energy density of gravitational radiation and its temperature. Thus, we obtain a purely macroscopic and classical equation of state for gravitational radiation. Based on this equation of state, we calculate the Gibbs energy, similar to what is observed for electromagnetic radiation. Consequently, we adopt a statistical model for the radiation and arrive at an intriguing result: a cubic dependence between a hypothetical gravitational quantum (graviton) and the frequency of the radiation. The article is divided as follows.  In section \ref{sec.2}, the Bondi-Sachs spacetime is described and the results of the gravitational energy and pressure of the radiation are given.  In section \ref{sec.3}, Boltzmann and Planck procedures are used for gravitational radiation.  With this, the energy of the graviton is established.  Finally, in the last section, the final considerations are presented.

\section{Bondi-Sachs Space-Time} \label{sec.2}

The purpose of this section is to briefly present the Bondi-Sachs space-time, in addition to the gravitational energy-momentum calculated within the framework of TEGR  in ref. \cite{malufBS}. It describes the gravitational radiation at spacelike and null infinities. The line-element, in the natural unities system, has the form

\begin{eqnarray}
ds^2&=&g_{00}\,du^2+g_{22}\,d\theta^2+g_{33}\,d\phi^2+2g_{01}\,du\,dr+2g_{02}\,du\,d\theta+2g_{03}\,du\,d\phi+
2g_{23}d\theta\,d\phi\,, 
\label{12}
\end{eqnarray}
where $u=t-r$ is the retarded time and the asymptotic behavior at $r\rightarrow \infty$ of the metric tensor is given by

\begin{eqnarray}
g_{00}&\simeq & -1+{{2M}\over r}\,, \nonumber \\
g_{01}&\simeq & -1+{{c^2+d^2}\over {2r^2}}\,,\nonumber \\
g_{02}&\simeq & l+{1\over r}(2cl +2d\bar{l} -p)\,, \nonumber \\
g_{03}&\simeq & \bar{l}\sin\theta + 
{1\over r}(-2c\bar{l}+2dl -\bar{p})\sin\theta\,, \nonumber \\
g_{22}&\simeq & r^2+ 2cr+2(c^2+d^2)\,, \nonumber \\
g_{33}&\simeq & \lbrack r^2- 2cr+2(c^2+d^2)\rbrack\sin^2\theta\,, \nonumber \\
g_{23}&\simeq & 2dr\sin\theta+{{4d^3}\over {3r}} \sin\theta\,,
\label{15}
\end{eqnarray}
with
$$l=\partial_2 c+2c\,\cot\theta+\partial_3 d\,\csc \theta\,,$$ and
$$\bar{l}=\partial_2 d +2d\,\cot\theta -\partial_3 c\csc \theta\,.$$ Here $M(u,\theta )$, $\partial_0 c(u,\theta)=\partial c/\partial u$ and $\partial_0 d(u,\theta) $ are the mass aspect, the first and second news functions respectively. The functions $p$ and $\bar{p}$ are defined in references of \cite{malufBS}. They do not contribute to the energy-momentum and therefore will not be detailed here. The time derivative of the mass aspect yields the loss of mass, while the news functions are interpreted as degrees of freedom of gravitational radiation. In the TEGR the energy-momentum depends on the choice of the tetrad field, which represents the choice of the observer.  For that, a set of tetrads was chosen that satisfies the condition $e_{(0)}\,^i=0$. With the help of the tetrad field, the Weitzenb\"ock connection can be defined, i.e.

$$\Gamma_{\mu\lambda\nu}=e^{a}\,_{\mu}\partial_{\lambda}e_{a\nu}\,,$$
whose antisymmetric part determines the following torsion tensor

\begin{equation}
T^{a}\,_{\lambda\nu}=\partial_{\lambda} e^{a}\,_{\nu}-\partial_{\nu}
e^{a}\,_{\lambda}\,. \label{4}
\end{equation}
It is worth noting that the Weitzenb\"ock connection relates to the Christoffel symbols by the following mathematical identity

\begin{equation}
\Gamma_{\mu \lambda\nu}= {}^0\Gamma_{\mu \lambda\nu}+ K_{\mu
\lambda\nu}\,, \label{2}
\end{equation}
where
\begin{eqnarray}
K_{\mu\lambda\nu}&=&\frac{1}{2}(T_{\lambda\mu\nu}+T_{\nu\lambda\mu}+T_{\mu\lambda\nu})\,,\label{3}
\end{eqnarray}
is the contorsion tensor. Such an identity allows writing the curvature scalar in terms of torsions, i.e.

\begin{equation}
eR(e)\equiv -e(\frac{1}{4}T^{abc}T_{abc}+\frac{1}{2}T^{abc}T_{bac}-T^aT_a)+2\partial_\mu(eT^\mu)\,,\label{eq5}
\end{equation}
where $e=det(e^a\,_\mu)$. It should be noted that the left-hand side of the above expression is the Hilbert-Einstein Lagrangian density, thus the TEGR Lagrangian density takes the following form

\begin{eqnarray}
\mathfrak{L}(e_{a\mu})&=& -\kappa\,e\,(\frac{1}{4}T^{abc}T_{abc}+
\frac{1}{2} T^{abc}T_{bac} -T^aT_a) -\mathfrak{L}_M\nonumber \\
&\equiv&-\kappa\,e \Sigma^{abc}T_{abc} -\mathfrak{L}_M\;, \label{6}
\end{eqnarray}
where $\mathfrak{L}_M$ is the Lagrangian density of matter fields, $\kappa=\frac{1}{16\,\pi}$ is the coupling constant in natural units and $\Sigma^{abc}$ is defined by

\begin{equation}
\Sigma^{abc}=\frac{1}{4} (T^{abc}+T^{bac}-T^{cab}) +\frac{1}{2}(
\eta^{ac}T^b-\eta^{ab}T^c)\;. \label{7}
\end{equation}
The derivative of this Lagrangian density with respect to the tetrad leads to the following field equation

\begin{equation}
\partial_\nu\left(e\Sigma^{a\lambda\nu}\right)=\frac{1}{4\kappa}
e\, e^a\,_\mu( t^{\lambda \mu} + T^{\lambda \mu})\;, \label{10}
\end{equation}
where

\begin{equation}
t^{\lambda \mu}=\kappa\left[4\,\Sigma^{bc\lambda}T_{bc}\,^\mu- g^{\lambda
\mu}\, \Sigma^{abc}T_{abc}\right]\,, \label{11}
\end{equation}
which is interpreted as the energy-momentum of the gravitational field. Such a definition has been shown to be quite consistent over the years. Hence the energy-momentum vector is defined by

\begin{equation}
P^a = \int_V d^3x \,e\,e^a\,_\mu(t^{0\mu}+ T^{0\mu})\,. \label{14}
\end{equation}
It is important to realize that the energy-momentum vector is invariant under coordinate transformations and is covariant under Lorentz transformations. This means that the zero component, identified with the energy, has all the qualities expected for a consistent definition of gravitational energy.

Thus, for the Bondi-Sachs metric, the energy reads \cite{malufBS}

\begin{equation}
P^{(0)}=4\kappa\int_0^{2\pi} d\phi \int_0^{\pi}d\theta \sin\theta
\biggl[M+\partial_0 F\biggr]\,, 
\label{25}
\end{equation}
where

\begin{equation}
F=-{1\over 4}\biggl(l^2 + \bar{l}^2 \biggr) +{1\over 2}c^2 +d^2
\label{26}
\end{equation}
It is worth noting that the term $\partial_0 F$ generalises the standard Bondi-Sachs energy. The momentum is given by

\begin{eqnarray}
P^{(i)}&=& 4\kappa\int_0^{2\pi}d\phi \int_0^{\pi}d\theta \sin\theta \biggl[
(M+\partial_0 F) \hat{r}^i \nonumber \\
&& +{1\over 4} (l\partial_0 M)\hat{\theta}^i +
{1\over 4}(\bar{l}\partial_0 M)\hat{\phi}^i \biggr]\,.
\label{31}
\end{eqnarray}
It should be noted that the energy-momentum is presented in terms of its components. The energy can also be expressed as 
\begin{equation}
P^{(0)}=\int d^3x \,\epsilon\,, 
\label{27}
\end{equation}
where $\epsilon =4\kappa\,\partial_r \biggl[M+\partial_0 F\biggr] $, assuming that a realistic physical system, like a radiating star, does not have a singularity. That means $\epsilon$ is the volumetric energy density. The derivative of the momentum with respect to time is precisely the force, thus
\begin{equation}
\frac{d P^{(i)}}{dt}=-\int dS_j \,\phi^{(i)j}\,, 
\label{28}
\end{equation}
where $\phi^{(1)1} =p_r=-4\kappa\,\frac{\partial}{\partial t} \biggl[M+\partial_0 F\biggr] $ is the radial pressure~\cite{maluf}. There is, therefore, a well-defined equation of state for this gravitational radiation 

\begin{equation}
\epsilon=p_r\,,
\end{equation}
that is, the energy density $\epsilon$ is equal to the radial pressure $p_r$.  Next, the consequences of this equation of state are analyzed.

\section{Thermodynamics of Gravitational Radiation} \label{sec.3}

The first law of thermodynamics is essentially a conservation law in which the heat produced by a heat engine minus the work done by the engine equals the internal energy of the system.  Although historically established in the context of heat engines, thermodynamics has a more fundamental character. It was this fundamental feature of thermodynamics that led Boltzmann to propose an ingenious interpretation of the entropy thermodynamic potential.  Until then, entropy accounted for the reversibility of thermodynamic phenomena.  Boltzmann proposed that it should be proportional to the logarithm of the number of possible states of the system.  The first law formulated in terms of entropy is then

\begin{equation}
TdS=dU+pdV\,,
\end{equation}
which in terms of the free energy $F=U-TS$ is

\begin{equation}
dF=-SdT-pdV\,.
\end{equation} 
It should be noted that the free energy is a thermodynamic potential which depends on the variables temperature and volume $F\equiv F(T,V)$, thus
$S=-\left(\frac{\partial F}{\partial T}\right)_V$ and $p=-\left(\frac{\partial F}{\partial V}\right)_T$. Hence $\left(\frac{\partial S}{\partial V}\right)_T= \left(\frac{\partial p}{\partial T}\right)_V $. Then the first law of thermodynamics reads

\begin{equation}
T\frac{dp}{dT}=\epsilon+p\,,\label{primeira}
\end{equation}
where $\epsilon=\frac{\partial U}{\partial V} $  and $p$ is the pressure. One should not confuse it with the quantity in the Bondi-Sachs space-time. It is worth noting that such an equation yields the Stefan-Boltzmann law for the electromagnetic radiation equation of state $p=\epsilon/3$.

Let's consider the equation of state for gravitational radiation obtained in the previous section $\epsilon=p_r$, we identify the thermodynamic energy $U$ with the gravitational energy $P^{(0)}$, therefore, equation (\ref{primeira}) reads

\begin{equation}
T\frac{dp_r}{dT}=2p_r\,,
\end{equation}
which yields the following dependence of pressure on the temperature

\begin{equation}
	p_r=\sigma \,T^2
\end{equation}
where $\sigma$ is the integration constant. The energy has the same dependence, $\epsilon= \sigma \,T^2$, while the entropy is $S=2\sigma TV$. It should be noted that entropy is calculated from the relation $\left(\frac{\partial S}{\partial V}\right)_T= \left(\frac{\partial p}{\partial T}\right)_V $. As an immediate consequence one can calculate the Gibbs free energy

\begin{eqnarray}
G&=&U-TS+PV \nonumber\\
&=&0\,.
\end{eqnarray}
This result is very interesting, as it also occurs for the electromagnetic radiation. In the latter case, the vanishing of $G$ implies the vanishing of the chemical potential $\mu$, since $G=\mu N$, where $N$ is the number of particles. Hence the number of particles must be non-zero (or there would be no physical system), which gave rise to the idea of photons for the electromagnetic field. This suggests that the gravitational radiation may have a quantum of energy $U_0$. Planck used Boltzmann's concept of entropy to derive the photon energy.  Here we're going to take a slightly different route to obtain the ``graviton'' energy. Thus, it is considered that gravitational radiation can be understood as a collection of discrete oscillators, each one with the energy $U_n=n U_0$.  Then the average value of the energy is

\begin{equation}
\bar{U}=\frac{\sum_n U_n \exp{\left(\frac{U_n}{kT} \right)}}{\sum_n\exp{\left(\frac{U_n}{kT} \right)}}\,,
\end{equation}
where $k$ is the Boltzmann constant. That yields
\begin{equation}
\bar{U}=\frac{U_0}{\exp{\left(\frac{U_0}{kT}\right)}-1}\,.
\end{equation}
The number of oscillators that exist between frequencies $\nu$ and $\nu+d\nu$ is $dN=8\pi V\nu^2d\nu$. That leads to the following energy density 

\begin{equation}
\epsilon= \frac{8\pi U_0\nu^2}{\exp{\left(\frac{U_0}{kT}\right)}-1}\,,
\end{equation}
which can be used to define the intensity of the radiation $I=\frac{\epsilon}{4\pi}$. Such a radiation intensity should reflect the squared temperature behavior when integrated over all frequencies, i.e.,
\begin{equation}
\int I(\nu , T)d\nu=\sigma\, T^2\,.
\label{27}
\end{equation}
If we assume that the quantum of energy of the gravitational radiation depends on the frequency of the radiation as a power law, then in view of the previous equation the energy is

\begin{equation}
U_0=\alpha\,\nu^3\,,
\end{equation}
which is the energy of the graviton. The equation (\ref{27}) may be rewritten as
$$\sigma T^2= \frac{2k^2T^2}{3c^2\alpha}\int_0^\infty \frac{x}{e^x-1}\,dx\,, $$ once the variable $x=\frac{\alpha \nu^3}{kT}$ is settled. Here SI units are used, in this system of units $c$ is the speed of light, which cannot be confused with the Bondi-Sachs space-time function. Since the integral above has a well-defined value $\int_0^\infty \frac{x}{e^x-1}\,dx =\frac{\pi^2}{6}$, then the $\sigma$ constant in terms of the Boltzmann constant and the $\alpha$ constant is given by
 $$\sigma= \frac{\pi^2k^2}{9c^2\alpha}\,.$$ In order to determine the value of such a constant, it should be noted that the alpha constant has a dimension of $J\,s^3$, the meaning of this result is possibly that the constant, which appears in the energy of the graviton, is identified with Planck's constant itself times the squared Planck time, i.e., $\alpha\sim 10^{-122} $. Thus the order of magnitude of this constant is $\sigma\sim 10^{60}$, in SI units. This may indicate that the $\alpha$ constant is completely independent of Planck's constant. Perhaps only experimental measurement can establish the value of this constant.

\section{Conclusion} \label{sec.5}

In this article we show that the definition of gravitational energy and pressure, in the context of Teleparallelism Equivalent to General Relativity, leads to a well-defined equation of state for the Bondi-Sachs space. Such an equation of state for gravitational radiation induces a specific dependence on temperature, that is, $p_r=\sigma \, T^2$. It is interesting to note that such an expression is a direct consequence of the first law of thermodynamics along with the existence of the equation of state. Since the Gibbs free energy is zero, we assume that the gravitational radiation also obeys the Planck's hypothesis.  Therefore we adjust the quantum of energy to reproduce the correct dependence on the temperature.  We obtain the energy of the graviton as $U_0=\alpha\,\nu^3$. It is worth noting that if an experiment were to confirm the dependence of gravitational energy density on temperature, this would render the metric approach to gravitation unfeasible. This is due to the fact that any dependence on temperature implies the existence of an equation of state involving energy and pressure. In this way the gravitational energy, whose existence is at least controversial in the context of general relativity, would have a physical existence. In Addition, the first law of thermodynamics, together with an equation of state, predicts a dependence of the pressure (or the energy) on the temperature. On the other hand, the non-existence of the concept of gravitational energy would contradict the first law of thermodynamics, if any dependence on the temperature of a gravitational radiation is verified. Thus measuring the temperature of a gravitational wave is a viable way to rule out the non-localizability of the gravitational energy. In the last 15 years, a series of measurements involving pulsars indicate the existence of background gravitational waves, which were not necessarily produced by black holes~\cite{nanograv1, nanograv2}. They can be legitimate gravitational radiation and it is necessary to investigate what relationship there is between such waves and thermal energy. Thus an analogous experiment can be carried out in order to correlate the frequency with the temperature of these waves.

\end{document}